\documentclass[prl,showpacs,superscriptaddress,twocolumn]{revtex4}

\usepackage{graphicx}
\usepackage{amsmath}
\usepackage{tabularx}
\usepackage{bm}
\usepackage{euscript}
\usepackage{epsfig}
\usepackage{graphicx}
\usepackage{color}
\usepackage{amsmath,amsfonts}
\usepackage{exscale}
\usepackage{amsbsy}

\def\be{\begin{equation}}
\def\ee{\end{equation}}
\def\ba{\begin{eqnarray}}
\def\ea{\end{eqnarray}}

\begin{document}

\title{Spin-memory effect and negative magnetoresistance in hopping conductivity}
\author{Oded Agam}
\affiliation{Physics Department, Columbia University, New York, NY 10027, USA}
\affiliation{The Racah Institute of Physics, The Hebrew University of Jerusalem, 91904, Israel}
\author{Igor L. Aleiner}
\affiliation{Physics Department, Columbia University, New York, NY 10027, USA}
\author{Boris Spivak}
\affiliation{Physics Department, University of Washington, WA 98195, USA}

\begin{abstract}
We propose a mechanism for negative isotropic magnetoresistance in the hopping regime. It results from a memory effect encrypted into spin correlations that are not taken into account by the conventional theory of hopping conductivity. The spin correlations are generated by the nonequilibrium electric currents and lead to the decrease of the conductivity. The application of the magnetic field destroys the correlations thus enhancing the conductance. This effect can occur even at magnetic fields as small as a few gauss.
\end{abstract}


\maketitle

In strongly disordered conductors, where electronic states are
localized, the conduction is due to phonon-assisted tunneling
between localized states \cite{ShklovskiiEfros}. The magnetoresistance
(MR) in such hopping regime is not well understood.  In many
insulators the relative magnitude of MR is significantly larger than
that of  metals \cite{Bergman}, and
its features are less universal. Experimental measurements in the
hopping regime showed both positive (see
\cite{ShklovskiiEfros,Shkl,B.I.Shklovskii} and references therein) and
negative MR \cite{Laiko,Jiang,Miliken,Frydman,Friedman,Wang,Mitin,Hong}. In some
materials more complicated behavior was observed: a giant MR that
changes its sign from positive to negative as the magnetic field
increases \cite{KravchenkoSimo,Kravchenko}.

The  mechanisms that were suggested for hopping MR
can be roughly divided into two classes: orbital related and spin related
mechanisms. The orbital mechanism is associated with the modification of the hopping
amplitude by the magnetic field and, depending on the model,
leads to  positive \cite{ShklovskiiEfros,Shkl} or negative \cite{Nguen,B.I.Shklovskii}
MR. The characteristic magnetic field in these cases,
$H^{*}{\cal A}\sim\Phi_{0}={\hbar c}/{e}$, corresponds to a flux quanta
threading the effective area, ${\cal A}$, explored by an electron during
the tunneling event. A distinctive feature of the orbital mechanism in two dimensional
films is its anisotropy with respect to the direction of the magnetic
field.

The spin mechanisms for positive MR are related either to the reduction of the
density of states with the increase of the magnetic field due to its
effect on doubly occupied states \cite{Kamimura}, or to a
possible reconstruction of the state of the system,
see discussion in Refs.~\cite{Kravchenko,Ioffe}.  Both mechanisms produce isotropic MR in films,
and the characteristic magnetic field, $\mu_B H^{*}\sim \min (T,J)$,
is obtained from the competition between the magnetic energy of the spin, $\mu_B H$
 ($\mu_B$ is the Bohr magneton), and either the temperature $T$ or the exchange energy between spins, $J$.

\begin{figure}
\begin{center}
\includegraphics[width=0.9\columnwidth]{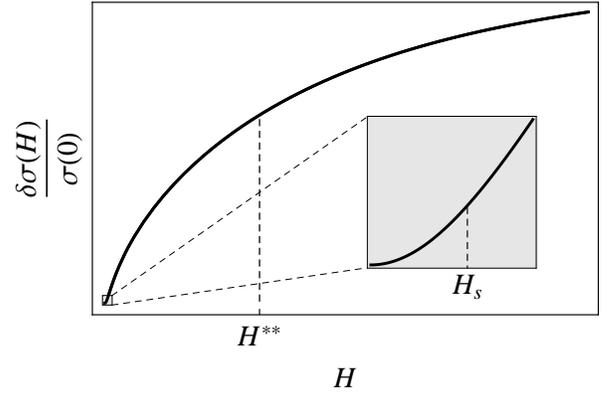}
 \caption{The positive magnetoconductance due to spin memory effect in
   the hopping regime. Here $\delta \sigma(H)$ is the difference
   between the conductance at finite magnetic field $H$ and at zero
   magnetic field $\sigma(0)$. The inset shows a magnified view of the
   curve near the origin. The characteristic fields $H^{**}$ and $H_{s}$ are determined by the hopping  and the spin relaxation times, respectively [ see Eq.~(\ref{Hs-def})].}
\label{fig0}
\end{center}
\end{figure}

The aforementioned theoretical studies  predict
rather high characteristic magnetic fields, $H^*$. Also the
experiments were mostly focused on relatively high
fields.

In this work we propose a spin-related mechanism for negative MR which takes place at weak magnetic fields sometimes as small as one gauss. It is isotropic and emerges from the long memory of nonequilibrium spin correlations created in course of electron transport. Our discussion will be mainly focused on the experimentally relevant regime where the characteristic hopping time, $\tau$, is much shorter than the spin relaxation time, $\tau_s$. In this regime the low magnetic field dependence of the conductance is demonstrated in Fig.~1, and is determined by two characteristic fields:
\be
H^{**}= \frac{1}{\delta g \mu_B\tau},~~~\mbox{and}~~~~H_s= \frac{1}{\delta g \mu_B\tau_s} ,  \label{Hs-def}
\ee
where $\delta g$ is the typical spatial fluctuation in $g$-factor. For the hopping conductivity, $\tau$ exponentially increases as the temperature decreases, therefore, in general $H^{**} \ll H^*$.

A qualitative explanation of the negative MR due to memory effect is
the following: Consider a situation where the hopping rate of an
electron between sites $i$ and $j$ depends on the relative spin
configuration of the hopping electrons and a spin located nearby at
site which we denote by $ij$ (see Fig.~2). In the presence of current
flowing through the system, a nonequilibrium correlation between the
spins at sites $i$ and $ij$ is created. For example, an electron
approaching  site $i$ from the bulk and making an unsuccessful attempt
to hop onto site $j$ will diffuse away and its spin density matrix
will depend on the spin at site $ij$. If it returns and attempts to hop one more time,
this attempt is not purely probabilistic. It is sensitive to the
previous history of the system, e.g. if the tunneling electron
formes a triplet state with the localized spin then it will still be in a
triplet state for the second attempt (for $H=0$), even though the tunneling between
the sites is inelastic. Let us now apply the magnetic field and neglect the
spin relaxation for a moment. If all the spins rotated in the same manner, the triplet state would always
remain triplet and there would be no magneto-resistance.
However, in  strongly disordered systems the
$g$-factor is random: This implies that the spins at different locations precess in
different manners  and therefore the spin correlations are destroyed by the magnetic field. The characteristic field where the MR saturates,
$H^{**}$, is obtained from the condition that the phase difference between the spins,
accumulated on a time scale of the order of the hopping time $\tau$, is of order one. Moreover, the return probability of an electron moving on the Miller-Abrahams network \cite{MA}
decays algebraically as a function of time. As a result the magnetoconductance exhibits a singular behavior
at small magnetic field. The spin relaxation introduces an upper
cutoff on the return time, and removes this singularity.

Although our mechanism is of general character \cite{chemical} (spin-dependent hopping rate is allowed by symmetry), to illustrate the effect we study a simple model where hops from site to site may
occasionally also involve a virtual transition through an occupied
state, as illustrated in Fig. 2.  We shall refer to the spin of the
electron on the occupied state as the ``link spin'' and denote it by
$\boldsymbol{s}_{ij}/2;\ \boldsymbol{s}_{ij}^2=3$.  The probability of
passing from $i$ to $j$ has an interference contribution of the direct transition and the indirect transition
which takes place when the link spin and the spin of the moving
electron form a singlet. We assume that the transition rate
associated with interference is small compared to the rate of
direct transition, and treat it to leading order in perturbation
theory. To further simplify the problem we assume that these link
spins are rare and that the concentration of electrons is very low,
namely the average occupation of each site (except the link spins) is
much smaller than one. We also neglect
the effects of the long range Coulomb interactions
\cite{ShklovskiiEfros}, which are crucial for the  temperature
dependence of the hopping transport but seem to be less important for
our mechanism of MR.

\begin{figure}
\begin{center}
\includegraphics[width=0.9\columnwidth]{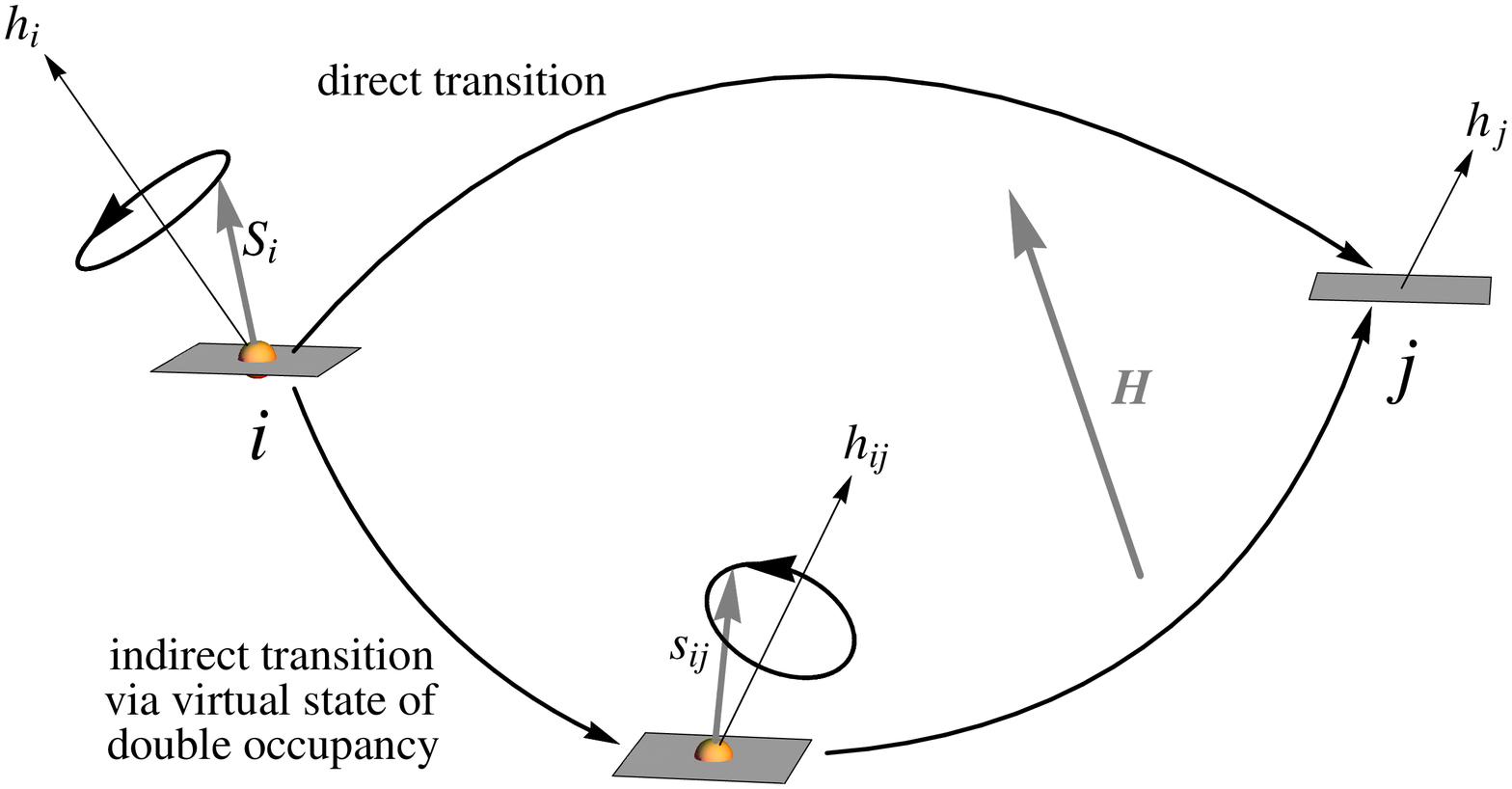}
\caption{A simplified hopping model in which electron may hop directly
  between two neighboring sites or go indirectly by forming a virtual
  singlet state with an electron at a nearby occupied site. The spins
  of the moving electron, ${\bf S}_i$, and the localized electron,
$\boldsymbol{s}_{ij}$, precess around different local fields, ${\bf h}_i$ and ${\bf h}_{ij}$, due to the spatial fluctuations of the $g$-factor.}
\label{fig1}
\end{center}
\end{figure}

 Let $P_i^0$, denote the probability of having no electron on site $i$ while if it is occupied by one electron its state is described by a $2\times 2$ matrix, $\hat{P}_i^1$, in the spin space. These quantities satisfy the normalization condition $P_{i}^{0}+ tr \hat{P}_{i}^{1} =1$
and therefore one may parametrize the state of an electron at site $i$ by $n_{i}= \mbox{Tr} \hat{P}_{i}^{1}$ and ${\bf S}_{i}= \mbox{Tr} \left( \boldsymbol{\sigma}\hat{P}_{i}^{1}\right)$ where $\boldsymbol{\sigma}$ are the Pauli matrices.

Average occupation numbers, $\langle n_i\rangle
$, are determined by
\begin{subequations}
\be
\label{n_i}
\begin{split}
\frac{d \langle n_i\rangle }{dt} & =
- \sum_{j}
\left[ \frac{ \langle n_i\rangle + \gamma_{ij}\left( \langle n_i\rangle - \langle {\bf S}_i\cdot \boldsymbol{s}_{ij}\rangle \right)}{\tau_{i \to j}}
-(i\leftrightarrow j)\right],
\\
\end{split}
\ee
where $1/\tau_{i \to j}$ denotes the bare transition rate from site
$i$ to site $j$. The first term on the right hand side accounts for
the decrease of the average occupation due to hop from site $i$ to
site $j$. It contains two contributions. The first $n_i/\tau_{i\to j}$
is associated with the direct transition, for simplicity we assumed
$n_k \ll 1$ and neglected the factor $1-n_k$. The effect of the
correlations in $\langle n_in_j\rangle$ was considered before
\cite{Richards77} and it does not change the MR. The second contribution,
proportional to $\gamma_{ij}\left( \langle n_i\rangle - \langle {\bf
    S}_i\cdot \boldsymbol{s}_{ij}\rangle \right)$, is associated with
the interference term of going through the virtual state [we define
$\boldsymbol{s}_{ij}\equiv\boldsymbol{s}_{ji},\ \gamma_{ij}\equiv \gamma_{ji}$]. This
transition occur only for the  moving electron and the
local spin forming a singlet. The corresponding contribution is
proportional to the small parameter of indirect transition
$|\gamma_{ij}|\ll 1$. The $(i\leftrightarrow j)$ term
describes the transition from site $j$ to site $i$.

The electron spin dynamics is described by
\be
\begin{split}
\frac{d \langle {\bf S}_i\rangle }{dt}& =  {\bf h}_i\times\langle {\bf S}_i \rangle
\\
&- \sum_{j}
\left[ \frac{ \langle {\bf S}_i\rangle + \gamma_{ij}\left( \langle {\bf
        S}_i\rangle - \langle n_i\boldsymbol{s}_{ij}\rangle \right)}{\tau_{i \to j}}
-(i\leftrightarrow j)\right],
\end{split}
\raisetag{50pt}
\label{si}
\ee
with the first term describing the spin precession [ ${\bf h}_i=  \hat{g}_i \mu_B  {\bf H}$ is local field acting on the
electron spin at site $i$, and $\hat{g}_i$ is the corresponding gyromagnetic
tensor], and the second line describes the same hopping processes as in
Eq.~(\ref{n_i}) (we neglect the direct effect of the magnetic field on
the hopping rates \cite{Kamimura}). Finally, the dynamics of the link spin is a pure precession:
\be
\frac{d \langle \boldsymbol{s}_{ij}\rangle }{dt} =  {\bf h}_{ij} \times \langle \boldsymbol{s}_{ij}\rangle,
\label{sij}
\ee
where
${\bf h}_{ij}=  \hat{g}_{ij} \mu_B  {\bf H}$. Exchange fields and the relaxation of the
spin via hopping involve terms of the order of $\gamma_{ij}^2$ which
we neglect. Other mechanisms of the spin relaxation will be included
later through the phenomenological relaxation time, $\tau_s$.
\label{dynamics}
\end{subequations}

The relation between Eqs.~(\ref{dynamics}) and the
corrections to Miller-Abrahams network can be understood as follows:
In equilibrium $ \langle {\bf S}_{j}\cdot
\boldsymbol{s}_{ij}\rangle=0$, and detailed balance implies
$n_{i}^{eq}/\tau_{i\to j}=n_{j}^{eq}/\tau_{j\to i}$, where $n_i^{eq}$
denotes the equilibrium occupation number. We redefine  the variables describing the nonequilibrium state of the system:
\begin{equation}
\langle n_{i} \rangle  \to n_{i}^{eq}(1+\psi_{i}), \quad {\bf S}_i
\to n_i^{eq} \tilde{\bf S}_i, \quad \boldsymbol{s}_{ij}
\to  \boldsymbol{s}_{ij}.
 \label{newvariables}
\end{equation}
  Equation~(\ref{n_i}) reduces  to
\begin{eqnarray}
n_i^{eq}\frac{ d \psi_i} {dt}= \sum_{j} \frac{1}{\tau_{ij}} \left[ \psi_j-\psi_i - \gamma_{ij} \left\langle \left( \tilde{\bf S}_j - \tilde{\bf S}_i \right)\cdot \boldsymbol{s}_{ij} \right\rangle  \right], \label{eqpsi}
\end{eqnarray}
where we defined
\begin{equation}
\frac{1}{\tau_{ij}} = \frac{n_i^{eq} (1+ \gamma_{ij})}{\tau_{i \to j}} = \frac{n_j^{eq} (1+ \gamma_{ij})}{\tau_{j \to i}}
\end{equation}
as the equilibrium transition rate between sites $i$ and $j$.
If there were no link spins, $\gamma_{ij}=0$, Eq~(\ref{eqpsi}) would
describe the Miller-Abrahams random resistor network
with the conductance of the link $i-j$ given by $ e^2/(T \tau_{ij})$.
The essence of the memory effect is that  symmetry allows
nonequilibrium spin correlations  to be a linear function of the
occupation numbers, i.e. assuming locality,
\begin{equation}
\gamma_{ij} \left\langle \left( \tilde{\bf S}_j - \tilde{\bf S}_i \right)\cdot \boldsymbol{s}_{ij} \right\rangle   = Q_{ij}(H) (\psi_j-\psi_i), \label{Qdef}
\end{equation}
where $Q_{ij}(H)$ is a function of the magnetic field, $H$. Therefore, as
follows  from (\ref{eqpsi}) and (\ref{Qdef}),  the conductances
are:
\begin{equation}
G_{ij} = \frac{e^2}{T \tau_{ij}} [1- Q_{ij}(H)]. \label{eq:conductance}
\end{equation}

In order to calculate the function $Q_{ij}(H)$, we need the equation
for the correlator
${\cal C}^{\alpha\beta}_{l;ij}=\langle \tilde{S}_l^\alpha s_{ij}^\beta\rangle$
(where $\alpha,\beta=x,y,z$ label components of the corresponding
vectors).
The easiest way to obtain the equation is to remove the
$\langle\dots\rangle$ in Eqs.~(\ref{dynamics}), multiply Eq.~(\ref{si}) by
${s}_{ij}^\beta$ and Eq.~(\ref{sij}) by $S_{i}^\alpha$, add the results and
average them again. According to Eq.~(\ref{sij}), even in nonequilibrium,
$\langle s^{\alpha}_{ij}s^{\beta}_{kl}\rangle =\delta_{ik}\delta_{jl}\delta_{\alpha\beta}$.
Then, to the leading
order in $\gamma_{ij}$, we obtain [in the variables (\ref{newvariables})]
\be
\begin{split}
& n_l^{eq}
\left[\left(\frac{d
  }{dt}+\frac{1}{\tau_s}\right)
{\cal C}^{\alpha\beta}_{l;ij}
- \epsilon_{\alpha\gamma\delta} h_l^\gamma
  {\cal C}^{\delta\beta}_{l;ij}
-\epsilon_{\beta\gamma\delta} h_{ij}^\gamma
{\cal C}^{\alpha\delta}_{l;ij}\right]
\\
&
= - \sum_{k\neq l}  \frac{
{\cal C}^{\alpha\beta}_{l;ij}-{\cal C}^{\alpha\beta}_{k;ij}
}{\tau_{lk}} +
 \gamma_{ij} \delta_{\alpha\beta}  (\delta_{il}-
 \delta_{jl})\frac{\psi_i-\psi_j}{\tau_{i j}},
\end{split}
\label{spin-spin}
\ee
with  $\epsilon_{\alpha \beta \gamma}$ being the antisymmetric tensor,
and repeated indices should be summed over. The relaxation time ${\tau_s}$,
introduced phenomenologically,
describes all the spin non-conserving processes.
The last term is the source of nonequilibrium spin correlations
 which after generation propagates by diffusion on the Miller-Abrahams
 network and precess in the spatially varying local field.

Equations (\ref{n_i}) and (\ref{spin-spin}) form the complete
description of the transport for a fixed realization of the relaxation
rates and local fields. To obtain the physical conductivity one needs
to average over such realizations. It is done using the percolation
theory approach to the hopping conductivity  \cite{ShklovskiiEfros} as
we describe below.

We notice, that if there were no randomness in the $\hat{g}_i$-tensors the relevant quantity ${\cal
  C}^{\alpha\alpha}_{l;ij}$ would not depend on the magnetic field at
all, as all spins rotate in the same manner. The correlation function  (\ref{spin-spin}) is affected only by the fluctuations of both ${\bf h}_i$  and ${\bf h}_{ij}$ (i.e. the averaged field may be subtracted).
The diffusing spin (index $l$) experiences fluctuating field because it hops  from site to site
 and, therefore, its accumulated rotation is proportional to the square root of
time. On the other hand the  field on the link ${\bf
  h}_{ij}$ remains stationary and its effect is linear in
time. Thus, we can substitute ${\bf h}_l\to 0$,   $n_l^{eq}{\bf
  h}_{ij}\to \bar{n}|\delta h_{ij}|\hat{\bf z} $,
$n_l^{eq}/\tau_s \to \bar{n}/\tau_s$ in Eq.~(\ref{spin-spin}),where $\bar{n}$ is the ensemble average of the equilibrium occupation numbers.
With this simplification, the function $Q_{ij}$ from
Eq.~(\ref{eq:conductance}) can be related to the properties of the
diffusion on the  same Miller-Abrahams network.
Consider the probability, ${\cal P}_{mm'}(t)$, to find the particle at the time $t$
on the site $m$ provided that at $t=0$ it was on $m^\prime$:
\be
\frac{\partial {\cal P}_{mm'}(t)}{\partial t} + \sum_{n}  \frac{
{\cal P}_{mm'}(t)-{\cal P}_{nm'}(t)
}{\tau_{mn}}=\delta(t)\delta_{mm'};
\label{diffusion}
\ee
Solving Eq.~(\ref{spin-spin}) for the stationary case, and substituting the result into
Eq.~(\ref{Qdef}),
we find
\be
\begin{split}
Q_{ij}(H)&
=\frac{(\gamma_{ij})^2}{\tau_{ij}}\sum_{l=-1}^1\int_0^\infty
dte^{-t\bar{n}( il\delta h+1/\tau_s)}
\Delta_{ij}(t);
\\
& \Delta_{ij}(t)={\cal P}_{ii}(t)+{\cal P}_{jj}(t)-{\cal P}_{ij}(t)-{\cal P}_{ji}(t).
\end{split}
\label{sol1}
\ee

Equation (\ref{sol1}) enables us to draw important conclusions about
the magnitude of the MR, the characteristic fields and
its asymptotic behavior. Indeed, $\Delta_{ij}(t)\simeq 2,\ t\lesssim
\tau_{ij}$, and, as we will see later,  $t\Delta_{ij}(t) \to 0$,
in the limit $t \to \infty$.
This means that the total magnitude of
the integral in Eq.~(\ref{sol1}) is determined by short time, and
the MR saturates at $\bar{n}\delta h\tau_{ij} \simeq
1$ [see Eq.~(\ref{Hs-def}) with $\tau=\tau_{ij}^{typical} \bar{n}$] thus $Q_{ij}(H\to\infty) \approx
Q_{ij}(0)/3$. As $\gamma_{ij}\ll 1$, the effect on each resistor is
small, so that one can always recalculate the change of the observable conductivity
in terms of the average change of the conductances of the percolation
network:
\be
\frac{\sigma(H\to \infty)-\sigma(0)}{\sigma(0)}\sim A=\rho\overline{\gamma_{ij}^2}. \label{main1}
\ee
where $\rho$ is the probability of having a link spin between two
sites on the percolation cluster and the overbar denotes ensemble averaging.

Our calculations relied on the assumption that the number of occupied
sites is small, $\rho \ll 1$, and that the amplitude for transition
through a virtual state is also small, $|\gamma_{ij}|\ll 1$, which
implied that $A\ll 1$. However, in general, these parameters need not
be small and both can be of order unity. In this case the magnitude of
the memory effect is also of order unity.

The actual value of the saturation magnetic field
$H^{**}$ from Eq.~(\ref{Hs-def}) strongly depends on the hopping time
and may be anomalously small.
Consider, e.g. a two dimensional sample with the resistance $R\sim
10^{9}\Omega$. Then, the typical hopping
rate is $\hbar/(T\tau)\simeq 10^{-6}$ [see Eq.~(\ref{eq:conductance})], and $\delta g \bar{n}\mu_BH^{*}/T
\simeq 10^{-6}$. Now estimating  $\delta g\sim
0.01$, we obtain $\mu_BH^{*}/T\gtrsim 10^{-4}$ which correspond to the
fields of the order of gauss at $T\simeq 1 K$.

Let us discuss the MR at $H<H^{**}$. The hopping rates
$\tau_{ij}$ are exponentially distributed and the observable
conductivity  and diffusion are determined by sites belonging
to the percolation cluster. Studies of anomalous
diffusion on the percolation cluster concluded that
\be
\Delta_{ij}(t)=\left(\frac{\tau}{t}\right)^{1+d_s/2},
\quad {t} \gtrsim \tau,
\label{andiffusion}
\ee
where $d_s$ is referred to as a spectral dimension of the percolation
cluster (see e.g. Ref.~\cite{percolation} for a review).
For spatial dimensions $d=2,3$, $d_s$ is close to $1.3$ \cite{percolation}.

Substituting Eq.~(\ref{andiffusion}) into Eq.~(\ref{main1}), we find
\be
\frac{\delta \sigma(H)}{A\sigma(0)} \sim -\Gamma\left(-\frac{d_s}{2}\right)
\sum_{l=-1}^{1}
\left[\left(\frac{ilH}{H^{**}}+\frac{\tau}{\tau_s}\right)^{\frac{d_s}{2}}- \left(\frac{\tau}{\tau_s}\right)^{\frac{d_s}{2}}\right],
\label{main}
\ee
where $\Gamma(x)$ is the gamma-function and only the singular dependence
for even $d_s$ is retained [$-\Gamma(-0.65)=3.9$].
The resulting magnetoconductance  is sketched in Fig.~\ref{fig0}.
[Strictly speaking, the correlation length of the percolation cluster
on the Miller-Abrahams network is infinite only in the limit $T\to 0$.
Taking into account the finite correlation radius introduces the new
value of the characteristic field below which one has to replace $d_s$ by
$d$.]  It is interesting to point out that for small
fluctuation of the $g$-factor, $H^{**}\to \infty$, and Eq.~(\ref{main})
predicts a {\em positive} MR via direct dependence of
the spin relaxation rate on the magnetic field
$\tau_s(H)/\tau_s(0)=1+(H/H_c)^2$, where $H_c$ is
determined by the correlation time of a spin relaxation process \cite{Abragam}.

To conclude, we considered a minimal model of the negative MR
due to memory effects in the hopping regime. Even though within our model,
the amplitude of
the effect is small, it has the strongest non-analytic magnetic field
dependence and the characteristic fields smaller than that for all the
other mechanisms considered in the literature. Further interesting
development may be in the direction of the more detailed study of the
variable range hopping regime where the number of the link spins
within the hopping length becomes large.
In this case, the memory
mechanism is expected to affect not only the preexponential factor of
the conductivity but the exponent itself resulting in giant memory MR.

We acknowledge useful discussions with A. Kapitulnik. This research has been
supported by the United States-Israel Binational Science Foundation (BSF) Grant No. 2012134 (O.A. and B.S.) and Simons
foundation (O.A. and I.A.).

\end{document}